\documentclass[preprint,review,12pt]{elsarticle}

\usepackage{amssymb}
\usepackage{amsmath,bm}
\usepackage{colortbl,booktabs}
\usepackage{tabularx}
\usepackage{threeparttable}
\usepackage{multicol,multirow}
\usepackage{subfigure}
\usepackage [font=footnotesize,labelfont=bf]{caption}
\usepackage{rotating}
\usepackage{tikz}
\usepackage{hyperref}

\definecolor{my_green}{RGB}{30,150,30}

\usepackage{color}

\journal{Elsevier}

\begin{document}

\begin{frontmatter}

	\title{Towards heterogeneous parallelism for SPHinXsys}

	\author{Xiangyu Hu\texorpdfstring{\corref{mycorrespondingauthor}}{}}
	\cortext[mycorrespondingauthor]{Corresponding author.}
	\ead{xiangyu.hu@tum.de}
	\author{Alberto Guarnieri}

	\address{School of Engineering and Design, Technical University of Munich\\
		85748 Garching, Germany}

	\begin{abstract}
Simulations based on particle methods, such as Smoothed Particle Hydrodynamics (SPH), are known to be computationally demanding. While such methods have for long been executed in parallel on multi-core CPUs, in recent years the increasing adoption of many-core accelerators, such as GPUs. However, hardware fragmentation and vendor-specific programming interfaces are still characterizing their market. Hence, support for various hardware configurations may easily lead to non-trivial and less maintainable implementations. To leverage over some higher-level specifications have become available recently, such as the SYCL programming standard, this work highlights the initial effort in adopting the SYCL standard for the execution of SPHinXsys, an open-source multi-physics library. The result is an execution model able to run the same implementation on variable (heterogeneous) hardware, with considerable speed-up compared to the current multi-core CPU parallelization. Among others, representation of data-structures for parallel access, communication strategies, and parallel methods for data sorting will be topics discussed in depth. Benchmarks has also been presented, showcasing performance comparisons between the current multi-core CPU implementation and the newly introduced SYCL parallelization with a GPU back-end.
	\end{abstract}

	\begin{keyword}
		Smoothed particle hydrodynamics  \sep GPU acceleration  \sep SYCL \sep data-structure
	\end{keyword}

\end{frontmatter}
%
%
\section{Introduction}
As smoothed particle hydrodynamics (SPH) is a typical particle-based method, 
many SPH libraries suffer from high computational costs 
due to intensive particle interactions. 
Compared to the multi-cores CPU system, 
the many-cores device (typically GPU) system
provides a much higher performance/cost ratio
for the intermediate (multi-million particles) scale simulations,
which are frequently encountered in industrial applications.
On the other hand, since the versatile functionality of CPU system,
it is expected that SPH libraries are using heterogeneous parallelism
so that they are able to take advantage of the both.

Currently, some SPH libraries already offer GPU support, 
either through vendor-specific implementations 
\cite{dualsphysics} \cite{sph-gpu} \cite{shallow-water-sph}, 
or heterogeneous programming using the OpenCL standard \cite{aquagpusph} \cite{pysph}.
An important issue of these implementations is 
that they require computing kernels to be defined,
as a low-level specification,
with specific directives separated from the rest of the code, 
rendering it less easily programmable and maintainable.

\section{Open-source multi-physics library SPHinXsys} 
SPHinXsys \cite{sphx}, 
is an open-source C++ SPH multi-physics simulations library.
It addresses the complexities of fluid dynamics, structural mechanics, 
fluid-structure interactions, thermal analysis, 
chemical reactions and AI-aware optimizations. 
SPHinXsys has offered CPU parallelism 
using Intel’s Threading Building Blocks (TBB) library \cite{tbb} 
and several features suitable for open-source based development. 

First, since the parallel execution is encapsulated into 
the low-level classes decoupled from the SPH method, 
the developers, as typical mechanical engineers,
only need to work the numerical discretization
without concerning parallelization.
Second, variable testing approaches, including unit test, google test 
and regression test, have been incorporated
so that refraction, adding new features and other modifications 
can be implemented without the worries about the already established functionalities.    
Third, automated cross-platform continuous integration/development (CI/CD) tests 
have being carried out on the open-source development 
platform frequently for any modification of 
the main branch of SPHinXsys repository.
In addition, SPHinXsys makes the uses of several third particle libraries,
such as Simbody library for multi-body dynamics, 
Pybind11 library for generating  python interface used 
for machine learning and optimization applications.
\section{SYCL Programming Standard}
Different from the other GPU-able SPH libraries, 
our work is based on the SYCL standard \cite{sycl} for parallel computing.
SYCL specification defines a new single-source
ISO C++ compliant standard with compute acceleration. 
It aims to be an high level abstraction that 
can be programmed as classical CPU code, 
without accelerator-specific directive 
and application interface (API) calls.
In this work, the Intel SYCL implementation,
i.e. Data Parallel C++ (DPC++) \cite{tbb}
is chosen, because it is a part of Intel’s oneAPI suit 
which also includes the TBB library already used in SPHinXsys.

A SYCL computing kernel 
(sharing the same term with the smoothing kernel used in SPH) 
is a scoped block of code executed on a SYCL device
under the global context \texttt{queue} and 
the local one \texttt{command\_group} 
initiated by the host.
Note that, as the SYCL host and device both can be CPU,
CI/CD jobs can test SYCL kernels on computers without GPU, 
such the standard runner provided by Github platform.
\section{Implementing SYCL kernels in SPHinXsys}
The long term aim for SPHinXsys development is 
employing unified codebase for the SPH methods
so the multi-physics simulation can be carried on 
heterogeneous computing systems in a collaborative way.
Therefore each SPH algorithm can be executed on host or device 
in sequential or parallel.

Before implementing SYCL kernels, 
SPHinXsys already has defined two execution policies,
namely \texttt{sequenced\_policy} and \texttt{parallel\_policy}
for CPU computing.
We deliberately implement the sequential execution on CPU
because it is essential for easy debug propose.
As SPHinXsys splits the execution,
namely \texttt{particle\_for} and \texttt{particle\_reduce}, 
and SPH methods, i.e. the classes inherited from the base \texttt{local\_dynamics}
with the same source code for both executions.

Our basic concept on implementing SYCL kernels is to extend 
the same programming pattern for CPU computing
by adding an extra execution policy,
namely \texttt{parallel\_device\_policy}  
to identify that the same source code of SPH methods 
will be executed in parallel on the device.

Though the concept is straightforward,
its implementation is not trivial since it relies 
on the an important assumption that the codes written in 
computing kernels should work for all execution policies. 
However, the source code of SPH methods previously used for CPU computing
generally are not executable on device
due to the widely used references, virtual functions 
and standard library dynamic memory allocations.
Therefore, a extended program pattern 
based on SYCL unified share memory (USM) is developed. 
The extended pattern has three layer, 
i.e. outer, kernel-shell and kernel, 
dependent on their distances to computing kernel 
which is the only one executed by the execution algorithms.

The outer layer is the same as before and defines the physical problem. 
Within this layer, the global variables are defined, 
the preprocess is done by generating particles respect to geometric information. 
Third-parties environment and methods, 
e.g. Simbody environment and methods, are referred directly. 
The kernel-shell layer is the SPH method definition layer,
which can be obtained from the CPU computing code with minimum modification. 
This layer is used to define individual SPH method. 
At this layer, the interface data structures, 
namely \texttt{discrete\_} and \texttt{singular\_variables}
are used to transfer the outer-layer data to those used in computing kernel. 
The kernel layer defines all the computing via kernels, 
and is obtained by splitting out the computing function from 
original method classes.
The objects of this layer are only instantiated and dispatched 
at the beginning of computing with 
the help of the kernel-shell layer interface and execution implementation. 
Note that the data transformed from kernel-shell layer has the form of raw pointer,
and can be accessed on host and device just as arrays.
\section{Cell linked-list, direct search and sorting}
Since the limitation of device hardware, 
SYCL specification does not allow the usage of \texttt{concurrent\_vector},
which is has been used in SPHinXys for constructing the cell linked-list for 
particle neighbor searching algorithms.
To achieve high performance on constructing cell linked-list 
on device concurrently,
atomic operations are used to avoid the possible thread conflicts. 

Again, due to the memory limit of GPU,
the full particle configuration, 
including neighbor list, kernel values 
used in SPhinXsys can not be used anymore. 
Therefore, direct search is used, 
i.e. the kernel values will be computed for all particles 
within the nearest cells in the computing kernels directly.

Similarly, 
due to poor performance of GPU on recursive algorithms,
the particle sorting executed by device uses \texttt{radix\_sort}
other than \texttt{quick\_sort}.
Note that, this is the only computing kernel that 
used different codes for device execution.  

\section{Performance Evaluation}
\begin{figure}[htbp]
	\centering
	\includegraphics[width=\textwidth]{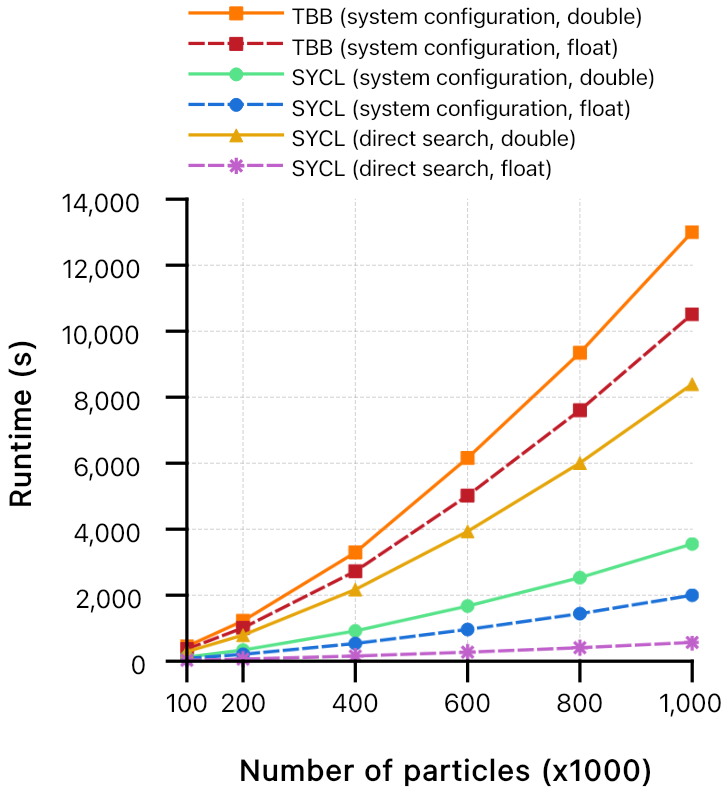}
	\caption{Simulation of 2-dimensional simulation of dam-break flow.}
	\label{fig:benchmarks-2d-dambreak}
\end{figure}

\begin{figure}[htbp]
	\centering
	\includegraphics[width=\textwidth]{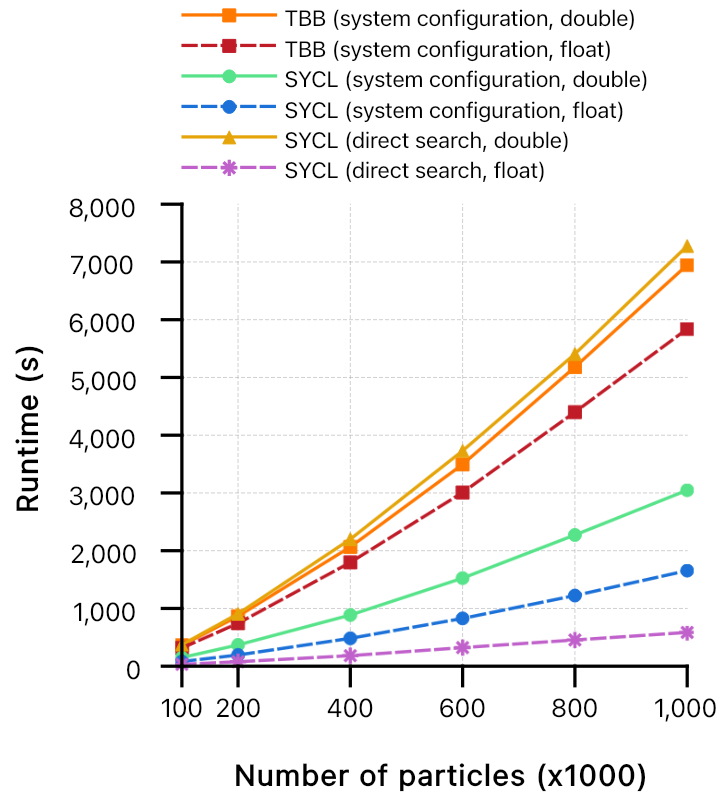}
	\caption{Simulation of 3-dimensional simulation of dam-break flow.}
	\label{fig:benchmarks-3d-dambreak}
\end{figure}

The dam-break flow simulation which very often  
has been used as reference is chosen as the benchmark test-case. 
Additionally, simulations have been executed with single
and double floating-point precision. 
Computations are carried out on a workstation 
composed of two Intel Xeon E5-2603 v4 and one NVIDIA GeForce RTX 2080 Ti.
Results presented in Figs.\ref{fig:benchmarks-2d-dambreak} and \ref{fig:benchmarks-3d-dambreak} 
delineate a performance improvement up to 27 times 
the execution on CPU using single precision. 
In particular, it is clear how single floating-point
precision on GPU runs considerably better 
than any other configuration when executed using direct search. 

\begin{figure}[htbp]
	\centering
	\includegraphics[width=\textwidth]{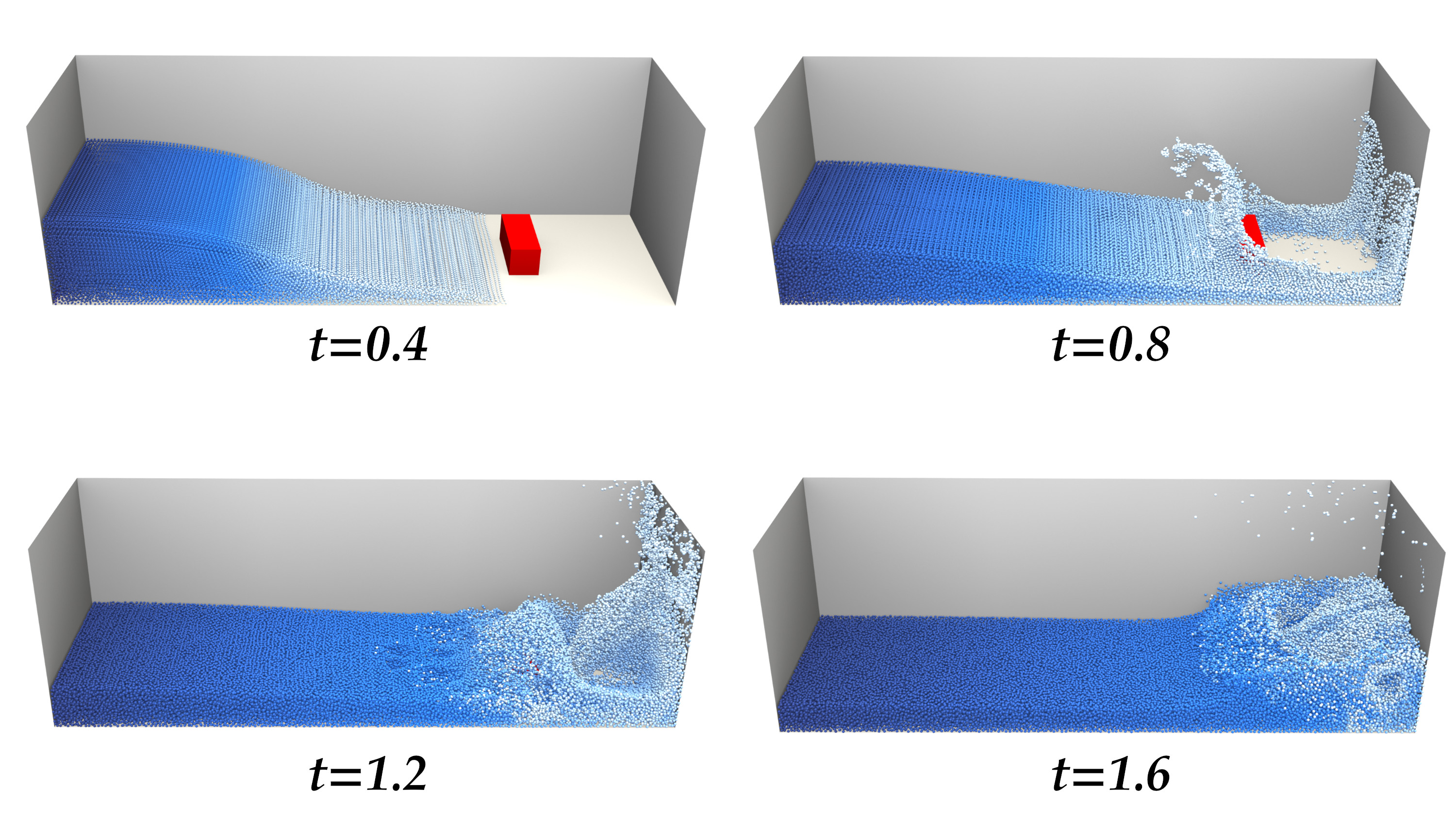}
	\caption{Snapshots for the dam-break with obstacle test-case.}
	\label{fig:kleefsman-snap-shots}
\end{figure}

In order to compare the achieved results with an existing GPU parallelization of SPH, 
SPHinXsys  has been tested against DualSPHysics, 
an SPH solver parallelized with CUDA.
The test-case defined by Kleefsman et al. in \cite{kleefsman}
is used to compare the two implementations.
The baseline results of DualSPHysics have been taken from an existing benchmark 
\cite{dualsphysics-2022} involving the same test-case. 
In particular, the results considered are simulation executed on an NVIDIA RTX 2080Ti.
The same test-case has been replicated in SPHinXsys and simulated on the same GPU.
As shown in Fig. \ref{fig:kleefsman-snap-shots}, the present simulation results are in quite good agreement with the literature. 
\begin{figure}[htbp]
	\centering
	\includegraphics[width=\textwidth]{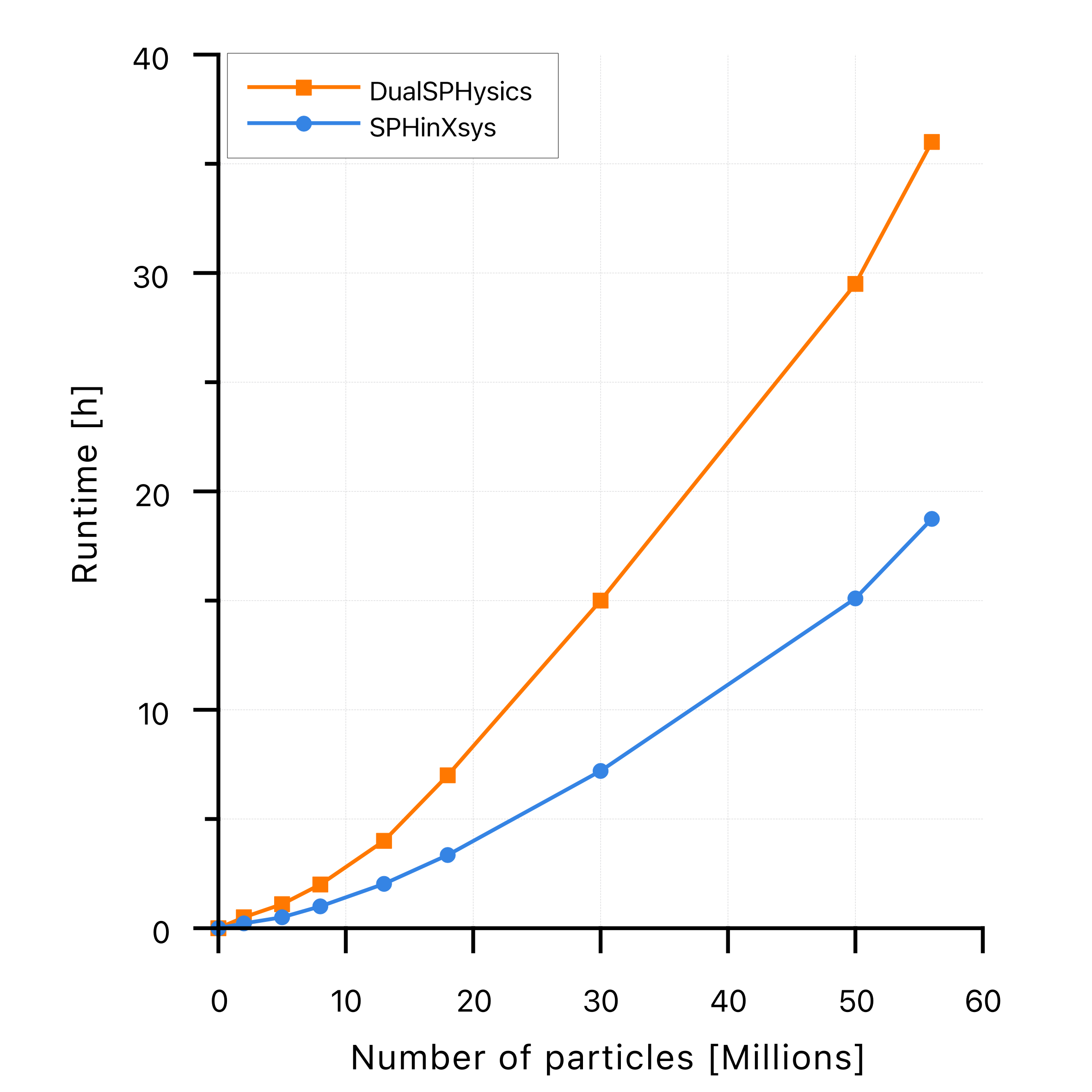}
	\caption{Runtime comparison between DualSPHysics and SPHinXsys for the dam-break with obstacle test-case.}
	\label{fig:kleefsman-runtime}
\end{figure}

\begin{figure}[htbp]
	\centering
	\includegraphics[width=\textwidth]{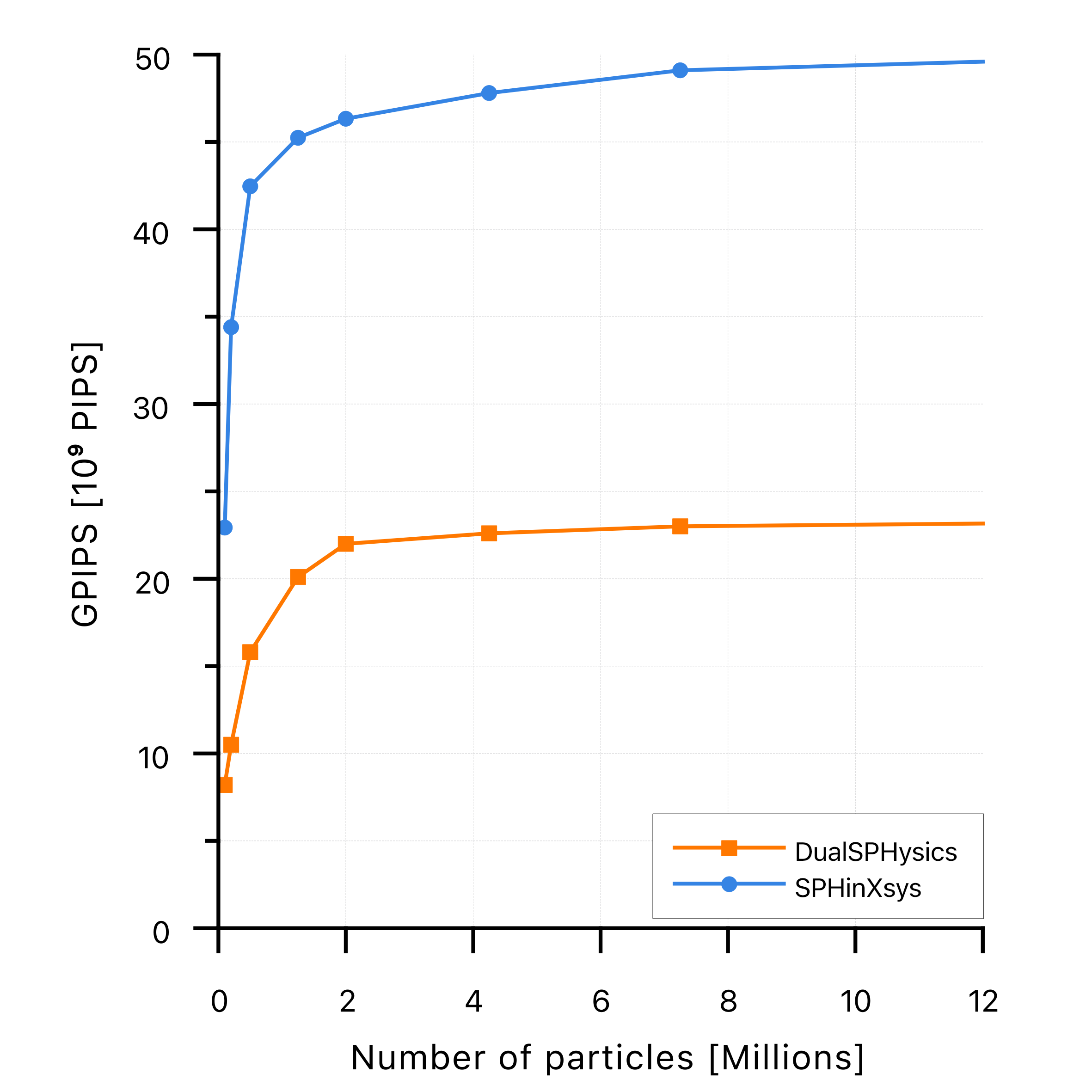}
	\caption{Comparison between DualSPHysics and SPHinXsys based on the number of particles interacting each second.}
	\label{fig:kleefsman-gpips}
\end{figure}
Results are presented in \autoref{fig:kleefsman-runtime}, and they show a SPHinXsys runtime that is half of the one needs by DualSPHysics.
Specifically, for the case with the maximum number of particles of 56 million, 
while DualSPHysics is able to computed the first two seconds of the simulation in 32 hours, 
SPHinXsys simulates the same period in 18 hours.
To confirm such results, 
Giga Particle Interactions Per Second (GPIPS) has also been computed, 
which is considered by Dom\'inguez et al. \cite{dualsphysics} 
a better unit of measurement compared to pure runtime.
The latter could in fact be affected by different timestep sizes, I/O output, etc.
It is found that SPHinXsys GPIPS are twice the amount of DualSPHysics, 
effectively confirming the runtime results, 
as shown in \autoref{fig:kleefsman-gpips}.
\section{Conclusions}
In this work, 
we presents the realization of heterogeneous parallelism for SPHinXsys 
based on the SYCL standard for GPU computing.
This is motivated by the expectation that,
besides accelerating the SPH simulation and 
easy programming and  maintaining, 
SPHinXsys is still able to maintain the already offered 
benefits for open-source based development
without notable compensation.

As the present implementation are 
actually more on generalizing the usage of data types
so that single computing kernel can be executed on 
host and device, sequenced or parallelized,
other than the exact numerical algorithm itself.
Heterogeneous parallelism is 
achieved with the minimum modification of 
the original program pattern.
This is especially suitable for 
the engineer developer as they do not need 
understand execution details of 
the numerical method they are developing.

Another advantage is that,
since generally only a single source code for the computing kernel 
to be executed with all execution policies,
SPHinXsys allows for the development and testing of numerical methods 
even in environments without GPUs or even DPC++ installed. 
If the source code for a computing kernel 
is crafted following our specified guidelines 
and prove functional in a standard platform, 
e.g. Linux system using the GNU compiler, 
they will seamlessly operate in environments 
equipped with DPC++ and GPU support.
\bibliographystyle{elsarticle-num}
\bibliography{literature}
%
%
\end{document}